# STUDY ON CODING TOOLS BEYOND AV1

*Xin Zhao*[*], *Liang Zhao*[*], *Madhu Krishnan*[*], *Yixin Du*[*], *Shan Liu*[*]
*Debargha Mukherjee*[†], *Yaowu Xu*[†], *Adrian Grange*[†]

[*]2747 Park Blvd, Palo Alto, CA 94306, United States, xinzzhao@tencent.com
[†]1600 Amphitheatre Parkway, Mountain View, CA 94043

**ABSTRACT**

The Alliance for Open Media has recently initiated coding tool exploration activities towards the next-generation video coding beyond AV1. With this regard, this paper presents a package of coding tools that have been investigated, implemented and tested on top of the codebase, known as libaom, which is used for the exploration of next-generation video compression tools. The proposed tools cover several technical areas based on a traditional hybrid video coding structure, including block partitioning, prediction, transform and loop filtering. The proposed coding tools are integrated as a package, and a combined coding gain over AV1 is demonstrated in this paper. Furthermore, to better understand the behavior of each tool, besides the combined coding gain, the tool-on and tool-off tests are also simulated and reported for each individual coding tool. Experimental results show that, compared to libaom, the proposed methods achieve an average 8.0% (up to 22.0%) overall BD-rate reduction for All Intra coding configuration a wide range of image and video content.

## 1. INTRODUCTION

In recent years, new generation of video coding standards have been developed by multiple international multimedia standardization organizations, including AOMedia Video 1 (AV1) [1] released in 2018 by Alliance for Open Media (AOMedia), AVS3 [2] released in 2020 by the Audio Video coding Standard (AVS) Workgroup of China and Versatile Video Coding (VVC) [3] standard released in 2020 by the Joint Video Experts Team (JVET) of ITU-T SG 16 WP 3 and ISO/IEC JTC1/SC29/WG11. These new-generation video coding standards have reportedly achieved substantial coding gain beyond the capability of their predecessors, including VP9, AVS2, and HEVC, and comparisons among the reference software developed for these standards have been conducted in the literature [4]. Looking forward, new requirement for next-generation video coding is driven by the emerging popularity of new video related applications (such as 8K, short music videos, cloud gaming and artificial intelligence in video), which are key technologies for video production and communications. With this regard, standardization organizations have started looking into the potential of developing new video coding methods beyond the capability of existing codecs. In ISO/IEC JTC 1/SC 29 WG1 (JPEG) and WG5 (JVET), expert groups have started exploration of neural network-based image and video coding technologies [5][6], respectively. In AOMedia, activities have been kicked off for the exploration of video coding technologies beyond AV1.

In this paper, several coding tools are investigated together as a package in the context of exploring new coding tools with capabilities beyond AV1. The coding tools are related to several key modules of traditional hybrid video coding structure, including block partitioning, intra prediction, transform and loop filtering. The investigated methods are described together with extensive experimental results and analysis.

The remainder of this paper is organized as follows. In Section 2, a brief review of AV1 coding tools is provided. In Section 3, the proposed tool investigations are categorized and described in different subsections. Experimental results, including both tool on/off tests are reported in Section 4 and the paper this concluded in Section 5.

## 2. BRIEF REVIEW OF AV1

Similar with other mainstream international video standards, AV1 also adopts the traditional hybrid video coding structure with extension and enhancement to different modules.

For block partitioning, in total 10 different partitioning patterns are supported, including binary-tree, quad-tree, T-type and 4-way horizontal/vertical splitting, and quad-tree splitting is the only partitioning pattern that supports further partitioning on each sub-block in a recursive manner.

For intra prediction, directional intra prediction modes have been extended from 8 (in VP9) to 56, and new non-directional intra prediction modes, including Paeth Predictor, recursive filtering modes, and CfL mode have been employed.

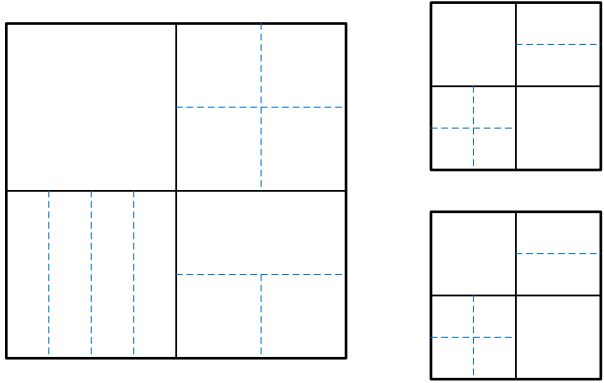

**Figure 1:** Illustration of coding block partitioning using SDP for luma (left) and chroma blocks (right).

For inter prediction, more flexible motion models with different number of control parameters (up to 6), e.g., local warped motion, are supported to simulate more complex motion beyond translational model. The motion vectors of non-adjacent neighboring blocks have been used for motion vector prediction. In addition, in AV1, an Overlapped Block Motion Compensation (OBMC) mode is also supported.

For transform coding, additional types of transform kernels (namely ADST, flipped ADST and IDT) are incorporated together with DCT-2 for performing the transform on residual samples. For entropy coding, a multi-symbol arithmetic coding engine is used to code a syntax with up to 16 different values. For coefficient coding, a bit-plane based coding approach is used.

For loop filtering, besides deblocking, AV1 also adopts the Constrained Directional Enhancement Filter and Loop Restoration (LR) filter. The LR filter further comprises two methods, i.e., Wiener filtering and self-guided filter. A film grain synthesis method has also been adopted in AV1 as a normative part of the codec to synthesize film grain.

For screen content coding, AV1 adopts the well-known Intra Block Copy (IBC) [7] and Palette mode [8] with different detailed designs compared to other existing video codecs.

## 3. CODING TOOLS BEYOND AV1

### 3.1. Partitioning

#### 3.1.1. Semi-decoupled partitioning

Image samples typically show strong correlations across different color channels since they represent same object, however, there can be still very different statistics for different color components. With YCbCr color space, luma component typically show finer details in textures whereas chroma component is usually much smoother. To utilize the correlation and adapt the different statistics among different color components, a semi-decoupled partition (SDP) method is proposed for coding block partitioning. With SDP, luma and chroma share the same coding block partitioning toward a specified partitioning depth. After this specified depth, different partitioning patterns can be chosen and signalled for luma and chroma components separately.

An example of block partitioning using SDP is shown in Fig. 1, where solid lines indicate partitioning boundaries that are shared and signalled together for luma and chroma, and dashed lines indicate partitioning boundaries that are signalled separately for luma and chroma. In this example, the luma and chroma components share one level of block partitioning and further splitting under this level is signalled separately. From experimental results, different image/video content and test conditions can have different preference on the shared depth of block partitioning. With this proposed SDP method, there is flexibility to decide how many levels luma and chroma components share the partitioning depth, and this is indicated in high-level syntax.

### 3.2. Prediction

#### 3.2.1. Improved intra mode coding

For intra prediction mode signalling, in AV1, 8 nominal intra prediction modes (IPMs) together with 5 non-directional IPMs are firstly signaled. Then, if current mode is directional IPMs, an offset is further signaled to indicate the angle delta relative to the associated nominal IPM. To better signal the intra prediction modes, intra prediction using adaptive prediction angles was proposed in [12], wherein only a subset of the IPMs are allowed and signaled for each block. The subset of IPM is adaptively selected according to the IPMs of neighboring blocks.

In addition, to capture the correlation between luma and chroma delta angles, it was proposed to derive the context of the chroma delta angles based on the delta angles of the co-located luma blocks [13]. In this paper, both methods are combined and implemented on top of SDP, wherein the top-left position in the chroma block is used to locate the corresponding luma block since luma and chroma blocks may have different partitions.

#### 3.2.2. Intra prediction using extended references

The idea of using multiple reference lines (MRL) for intra prediction was proposed in [10][11], and further investigated on top of AV1 in [12]. In this paper, further optimizations have been done when it is integrated together with other coding tools presented in this paper. For non-zero reference lines, instead of 2-tap bilinear interpolation, 4-tap edge preservation interpolation filter is used for generating intra prediction samples.

### 3.3. Transform

*3.3.1. Extension on primary transform*

Line graphs are mathematical structures consisting of sets of vertices and edges, which are used for modelling affinity relations between the objects of interest. In [19], separable LGTs are designed and optimized by learning line graphs from data to model underlying row and column-wise statistics of blocks residual signals, where the associated generalized graph Laplacian (GGL) matrices are used to derive LGTs.

In the proposed extension of primary transforms (EPT), the 4-point DST-VII in AV1 is replaced with a 4-point LGT with self-loop weights ($v_{c1}$, $v_{c2}$) = ($2w_c$, 0), which leads to a DST-IV. The principal basis of DCT-IV exhibits a steeper slope and thus can better characterize the statistics of residual samples in small block sizes. For larger blocks that involve 8-point and 16-point DST-IV in AV1, 8-point & 16-point LGT with self-loop weights ($v_{c1}$, $v_{c2}$) = ($1.5w_c$, 0) and ($v_{c1}$, $v_{c2}$) = ($w_c$, 0) are used, respectively. The 16-point LGT is essentially a DST-VII. All the LGTs described above (4-point, 8-point and 16-point) are implemented as matrix multiply. The LGT kernels are constructed by 8-bit integers that are tuned for orthogonality.

*3.3.2. Non-separable directional transforms*

To investigate the impact of introducing a non-separable transform scheme in AV1, a non-separable unified secondary transform (NUST) scheme was implemented, and results reported in [14]. In this paper, a non-separable directional transform (NSDT) method is proposed that further improves NUST.

Different from the method in [14], input to forward and inverse secondary transform are low frequency primary transform coefficients in zig-zag scan order instead of raster scan order. This helps to achieve more efficient decorrelation of neighboring low frequency coefficients.

In AV1, both intra and inter coded blocks can be further partitioned into multiple transform units with the partitioning depth up to 2 levels. In the proposed method, NSDT is limited to be applied to the root (depth 0) of the transform partitioning tree structure. A decrease in the software encoder run-time with minor impact in compression efficiency can be achieved by this change. In addition, the signaling of transform indices are further improved by using a prediction mode and transform block size dependent context for entropy coding the NSDT index.

### 3.4. Loop filtering

*3.4.1 Cross-component sample offset*

Cross-component video coding technology aims at capturing the statistical correlations across different color components, e.g., Cross-Component Linear Model, Joint Cb and Cr residual coding and Cross-Component Adaptive Loop Filter [20] applied in VVC. In this paper, a new non-linear loop filtering approach called Cross-Component Sample Offset (CCSO) is proposed. The proposed CCSO is featured by a non-linear offset mapping process implemented as a look-up-table (LUT). The filtering of CCSO utilizes a diamond shape filtering support that includes 5 samples as input.

In CCSO, given the collocated reconstructed luma sample $rl$ of the current chroma sample $rc$ that is to be filtered, the four surrounding samples of $rl$ are used as input, denoted as $p_0$, $p_1$, $p_2$, $p_3$. The delta values between $p_0$ - $p_3$ and $rl$ are further quantized into $d_0$ - $d_3$. A LUT is used taking $d_0$ - $d_3$ as input, and outputs an offset value $s$. Finally, the offset value $s$ is applied on $rc$. The LUT is optimized and signaled per Cb/Cr component. The enabling of CCSO is controlled at both frame- and block-level.

### 4. EXPERIMENTAL RESUTLS

The proposed coding tools have all been implemented on top of the research branch of libaom [16], the associated commit hash # is

```
994c889176bcabbc149445971ae12d1f0069eab9,
```

which corresponds to a recent version (released in Sep. 2020) of libaom research branch.

The test set has a total of 56 sequences, including 10 Class A1 (4K), 17 Class A2 (1080p), 8 Class A3 (720p), 6 Class A4 (360p), 4 Class A5 (240p) camera captured sequences and 11 Class B1 (synthetic) sequences. These test sequences are defined in the common test condition (CTC) specified by AOMedia Testing Subgroup [17]. The Bjøntegaard delta bitrate (BD-Rate) [18] is used to evaluate the coding gain. The quantization parameters (QP) settings are 23, 31, 39, 47, 55 and 63. The run-time complexity is measured by the ratio between the anchor and tested method, i.e., $T = T_{\text{Proposed}}/T_{\text{Anchor}}$. The BD-rates are calculated using several quality metrics, including luma PSNR/SSIM, chroma (Cb and Cr separately) PSNR/SSIM and overall PSNR/SSIM that is calculated using 6/8, 1/8 and 1/8 weightings on luma, Cb and Cr components quality scores, respectively.

The following encoder parameters are used for the simulations as defined in CTC. Please note that, Palette mode is also disabled (--enable-palette=0) because the current implementation of the proposed tools is not completely properly working with Palette mode yet at the moment of this paper submission.

```
All Intra (AI): --cpu-used=0 --passes=1 --end-usage=q
-cq-level=x --kf-min-dist=0 --kf-max-dist=0 --use-fixed-
qp-offsets=1  --limit=X  --deltaq-mode=0  --enable-tpl-
```

Table 2: Results of Tool-On tests for all proposed tools.

| Tools | All Intra | | | | | | Random Access | | | | | |
|---|---|---|---|---|---|---|---|---|---|---|---|---|
| | BDR-Y | BDR-U | BDR-V | YUV | $\Delta T_{Enc}$ | $\Delta T_{Dec}$ | BDR-Y | BDR-U | BDR-V | YUV | $\Delta T_{Enc}$ | $\Delta T_{Dec}$ |
| SDP | -1.3% | -11.7% | -12.9% | **-3.5%** | 85% | 127% | | | | | | |
| MRL | -1.4% | -0.9% | -1.1% | **-1.3%** | 166% | 104% | -1.1% | 0.3% | 0.9% | **-0.8%** | 129% | 99% |
| IMC | -1.8% | 1.7% | 1.8% | **-1.1%** | 160% | 106% | -0.9% | 1.4% | 1.9% | **-0.5%** | 112% | 100% |
| EPT | -0.2% | 0.2% | -0.1% | **-0.2%** | 210% | 117% | -0.1% | 0.1% | 0.1% | **-0.1%** | 131% | 108% |
| NSDT | -2.8% | 1.7% | 1.3% | **-2.0%** | 156% | 110% | -1.0% | 0.0% | 0.3% | **-0.8%** | 109% | 103% |
| CCSO | 0.3% | -4.9% | -5.7% | **-0.8%** | 117% | 120% | 0.4% | -5.4% | -5.9% | **-0.8%** | 108% | 114% |

Table 1: Results of Tool-Off tests for all proposed tools.

| Tools | All Intra | | | | | | Random Access | | | | | |
|---|---|---|---|---|---|---|---|---|---|---|---|---|
| | BDR-Y | BDR-U | BDR-V | YUV | $\Delta T_{Enc}$ | $\Delta T_{Dec}$ | BDR-Y | BDR-U | BDR-V | YUV | $\Delta T_{Enc}$ | $\Delta T_{Dec}$ |
| SDP | -1.1% | -13.8% | -14.6% | **-3.7%** | 89% | 113% | | | | | | |
| MRL | -0.5% | -0.5% | -0.5% | **-0.5%** | 86% | 100% | -0.2% | -0.2% | 0.0% | **-0.2%** | 107% | 103% |
| IMC | -0.6% | 0.1% | 0.1% | **-0.5%** | 93% | 104% | -0.2% | 0.6% | 0.6% | **0.0%** | 91% | 104% |
| EPT | -0.2% | 0.0% | -0.1% | **-0.2%** | 156% | 111% | -0.1% | 0.0% | 0.1% | **-0.1%** | 139% | 106% |
| NSDT | -2.8% | -0.5% | -0.7% | **-2.4%** | 130% | 103% | -1.1% | 1.1% | 1.5% | **-0.7%** | 122% | 104% |
| CCSO | 0.3% | -3.5% | -4.0% | **-0.5%** | 105% | 115% | 0.4% | -5.8% | -6.1% | **-0.9%** | 105% | 117% |

```
mode=0 --end-usage=q --enable-keyframe-filtering=0 --obu
--enable-palette=0

  Random Access (RA): --cpu-used=0 --passes=1 --lag-in-
frames=19 --auto-alt-ref=1 --min-gf-interval=16 --max-
gf-interval=16    --gf-min-pyr-height=4    --gf-max-pyr-
height=4 --limit=130 --kf-min-dist=65 --kf-max-dist=65 -
-use-fixed-qp-offsets=1   --deltaq-mode=0   --enable-tpl-
model=0 --end-usage=q --cq-level=<qp> --enable-keyframe-
filtering=0 --obu --enable-palette=0
```

The coding tools are tested using tool-on and tool-off tests. In a tool-on test, the anchor is libaom (AV1 compatible), and the test is the anchor with a specified individual coding tool being enabled. In a tool-off test, the anchor is libaom with all the proposed tool being enabled, and the test is the anchor with a specified individual coding tool being disabled. If the tool-on gain is more than tool-off gain, it means the coding gain of a specific coding tool overlaps other coding tools. If the tool-on gain is equal to or less than the tool-off loss, it means the coding gain of a specific coding tool is additive or even more efficient with the presence of other coding tools. In the following subsections, detailed test results will be discussed and analyzed.

### 4.1. Tool-on test

In this test, the coding performance of each individual tool is studied on top of AV1 without the presence of other proposed coding tools. For this investigation, the anchor is libaom research branch (hash tag #994c88*), and test is anchor with Tool X being enabled, where Tool X represent one of the proposed tools.

The results of Tool-On tests are tabulated in Table 2. From the results, it can be observed that the coding tools mainly contributing luma BD-rate gains are MRL, IMC and NSST, and the coding tools mainly contributing chroma BD-rate gains are SDP and CCSO. There is a significant 15% encoder run-time saving from SDP that comes from using larger transform sizes for chroma. The decoder run-time increase of SDP is mainly caused by implementation issue, which will be further optimized as future work. Most of the coding tools contribute only intra coding gains except for CCSO that operates on both key frame and inter frame.

### 4.2. Tool-off test

In this test, the coding performance of each individual tool is studied on top of AV1 with the presence of all other proposed coding tools. For this investigation, the anchor is libaom research branch (hash tag #994c88*) with all proposed coding tools except for Tool X being enabled, where Tool X represent one of the proposed tools. The test is libaom research branch with all proposed coding tools being enabled.

The results of Tool-Off tests are tabulated in Table 1. Comparing to the Tool-On test results, it can be observed that, among the investigated coding tools, the coding gain of MRL and IMC overlaps with other tools, while SDP, EPT, NSDT and CCSO provide relatively consistent coding gain regardless the presence of other coding tools. However, it is noted that the encoder run-time impact of MRL and IMC is also largely reduced when other coding tools are enabled, therefore, there are some encoder designs to be further studied and optimized in future.

**Table 3**: Results of proposed tool package for All-Intra configurations.

| Class | PSNR | | | | SSIM | | | | SIMD on | | SIMD off | |
|---|---|---|---|---|---|---|---|---|---|---|---|---|
| | BDR-Y | BDR-U | BDR-V | YUV | BDR-Y | BDR-U | BDR-V | YUV | $\Delta T_{Enc}$ | $\Delta T_{Dec}$ | $\Delta T_{Enc}$ | $\Delta T_{Dec}$ |
| Class A1 (4K) | -4.8% | -11.6% | -13.1% | -6.4% | -6.0% | -12.1% | -12.4% | -6.7% | 328% | 173% | 157% | 142% |
| Class A2 (2K) | -6.6% | -12.3% | -14.7% | -7.9% | -8.2% | -13.3% | -12.2% | -8.8% | 379% | 180% | 158% | 141% |
| Class A3 (720P) | -5.3% | -17.3% | -19.6% | -7.8% | -6.1% | -12.4% | -11.8% | -6.4% | 446% | 184% | 187% | 150% |
| Class A4 (360P) | -6.2% | -26.4% | -32.7% | -10.5% | -8.0% | -19.7% | -17.9% | -10.0% | 420% | 169% | 176% | 135% |
| Class A5 (240P) | -4.4% | -19.3% | -13.9% | -7.0% | -5.3% | -5.7% | 5.5% | -4.9% | 425% | 159% | 188% | 128% |
| Class B1 (Synthetic) | -6.8% | -17.1% | -16.3% | -8.8% | -7.1% | -18.0% | -15.6% | -8.8% | 303% | 171% | 157% | 141% |
| **Min** | | | | -22.0% | | | | -22.3% | 126% | 123% | 108% | 99% |
| **Max** | | | | -2.9% | | | | -2.4% | 614% | 259% | 245% | 221% |
| **Average** | **-5.9%** | **-15.8%** | **-17.3%** | **-8.0%** | **-7.1%** | **-14.0%** | **-12.2%** | **-7.9%** | **385%** | **176%** | **167%** | **142%** |

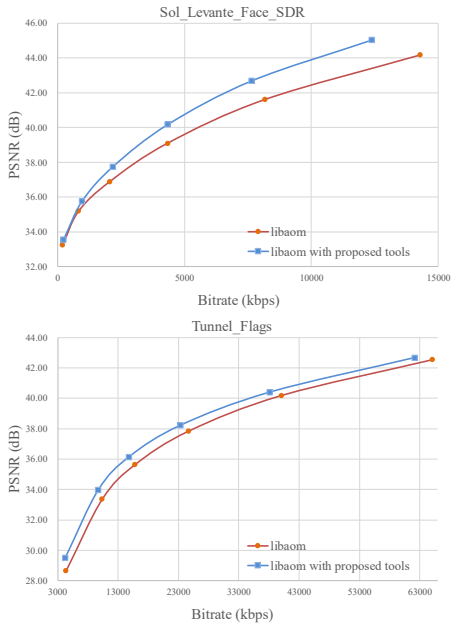

**Figure 2:** Illustration of coding block partitioning using SDP for luma (left) and chroma blocks (right).

### 4.3. Tool package

In this test, the combined coding performance of all proposed coding tools as a package is studied on top of AV1. The anchor is libaom research branch (hash tag #994c88*), and the test is libaom research branch with all proposed coding tools being enabled. The detailed results are listed in Table 3 for each class, using PSNR and SSIM as the metrics.

As shown in Table 3, the average coding gain is -8.0% and -7.9% when using PSNR and SSIM as the quality metrics, respectively. The coding gain can go up to 22% (for one sequence in Class B1). The encoder and decoder run-time is 385% and 176%, respectively, however, this is not fair comparison, since for directional intra prediction modes and 4-p, 8-p and 16-p primary transform using LGT, pure C code implementation is used while the anchor (libaom) is using SIMD optimization for these functions. In the last two columns of Table 3, we also compared the run-time when the related SIMD optimizations are turned off. The PSNR-bitrate curve for two sequences are shown in Figure 2.

### 4.3. Visual quality

Subjective visual quality of the image reconstructed by AV1 and proposed codec is also compared. As shown in Figure 3, the partial reconstructed picture of frame 14 of TunnelFlag (1080p) is shown. The picture coded by AV1 is using 20660 Bytes with QIndex 224, and the picture coded by proposed methods is using 20550 Bytes with QIndex 220. It is observed that reconstructed picture using proposed method show better visual quality, especially around the edges. The objective metric also shows 0.83dB, 1.11dB and 1.69dB higher PSNR for Y, Cb and Cr components, respectively. There are also other aspects that show subjective quality improvements, such as less color inconsistency, finer texture details, less distorted straight lines.

### 5. CONCLUSIONS

In this paper, several coding tools beyond AV1 are proposed, which improve coding efficiency based on a hybrid video coding structure, including block partitioning, intra prediction, transform and loop filtering. The coding tools are implemented together as a package to investigate the combined coding gain as well as individual coding gain when other coding tools are present. With the design described in this paper, a combined 8.0% coding gain for intra coding has been demonstrated using the proposed methods.

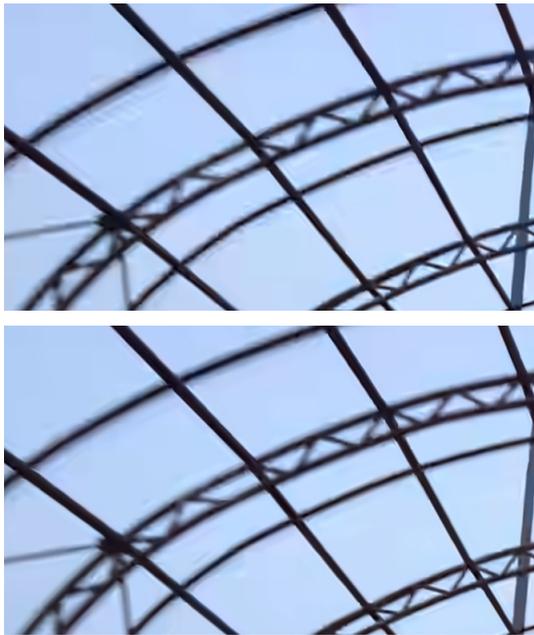

**Figure 3:** Comparison on the reconstructed image of TunnelFlag sequence using AV1 (top, 20660 Bytes) and proposed codec (bottom, 20550 Bytes).